\documentclass[11pt]{article}
\usepackage{epsf,epsfig,float,amssymb,latexsym,amsmath,amsthm,fancyhdr,pst-all}
\usepackage{graphics,psfrag}
\newcommand{\xpc}{\%}
\newcommand{\om}{\Omega}
\newcommand{\la}{\langle}
\newcommand{\ra}{\rangle}

\newcommand{\be}{\begin{equation}}
\newcommand{\ee}{\end{equation}}
\newcommand{\ba}{\begin{eqnarray}}
\newcommand{\ea}{\end{eqnarray}}

\newcommand{\nn}{\nonumber}

\newcommand{\vs}{\vspace{-0.20cm}}
\begin{document}

\thispagestyle{empty}

\vspace{2cm}

\begin{center}
{\Large{\bf Improved dispersion relations for $\gamma\gamma\to \pi^0\pi^0$}}
\end{center}
\vspace{.5cm}

\begin{center}
{\large Jos\'e A. Oller$^a$, Luis Roca$^a$ and Carlos Schat$^{a,b}$ }
\end{center}

\begin{center}
{\it {\it $^a$Departamento de F\'{\i}sica. Universidad de Murcia. E-30071,
Murcia. Spain.\\
$^b$ CONICET and
Departamento de F\'{\i}sica, FCEyN, Universidad de Buenos Aires, \\Ciudad Universitaria,
Pab.1, (1428) Buenos Aires, Argentina.\\
{\small oller@um.es~,~luisroca@um.es~,~schat@df.uba.ar}}}
\end{center}
\vspace{1cm}

\begin{abstract}
\noindent
We perform a dispersive theoretical study
 of the reaction $\gamma\gamma\to \pi^0\pi^0$ emphasizing
 the low energy region. The  large  
source of theoretical uncertainty  to calculate  
  the $\gamma\gamma\to\pi^0\pi^0$ total cross section for 
 $\sqrt{s}\gtrsim 0.5$~GeV within the dispersive approach is removed. This is accomplished 
 by taking  one more subtraction in the dispersion relations, where 
 the extra subtraction constant is fixed 
by considering new low energy constraints, 
one of them further refined by taking into consideration the $f_0(980)$ region. 
 This allows us to make sharper predictions for the 
cross section for $\sqrt{s}\lesssim 0.8$~GeV, below the onset of D-wave
contributions. In this way, were new more precise data on $\gamma\gamma\to\pi^0\pi^0$
 available  one might then distinguish between different parameterizations of the
 $\pi\pi$ isoscalar S-wave.  
We also elaborate on the width of the $\sigma$ resonance to $\gamma\gamma$ and provide 
new values.  
\end{abstract}

\vspace{2cm}

%\begin{center}
%Keywords:
%\end{center}

\newpage

%-----------------------------------------------------------------------------
\section{Introduction}
\label{sec:intro}
\def\theequation{\arabic{section}.\arabic{equation}}
\setcounter{equation}{0}

The reaction $\gamma\gamma\to\pi^0\pi^0$ measured in ref.\cite{crystal} offers the 
interesting prospects of having  a two-body hadronic final state and 
the important role of final state interactions in S-wave enhanced 
due to the null charge of the $\pi^0$. These two facts make this process very suited 
for learning about the non-trivial  $\pi\pi$ isospin ($I$) 0 S-wave. 
In addition, it was taken as an especially appropriate  
 ground test for Chiral Perturbation Theory ($\chi$PT) \cite{wein,glchpt}, 
since at lowest order this process  is zero and
at next-to-leading order  (one loop) is a prediction free of any
counterterm \cite{bc88,dhl87}. However, the one loop $\chi$PT prediction 
departs   very rapidly from data just above the threshold    
 and only the order of magnitude was rightly
foreseen. A two loop calculation in
ref.\cite{bgs94,gis05} was then undertaken with better  agreement with data \cite{crystal}.
 The three counterterms that appear at
${\cal O}(p^6)$  are fixed by the resonance saturation hypothesis.
Other approaches supplying higher orders to one loop $\chi PT$  by
taking into account unitarity and analyticity  followed \cite{dh93,dp93,oo98}.
 Ref.\cite{oo98} is a Unitary $\chi$PT  calculation in production processes
\cite{oo98,phi1,hyper,sfm,vfm,anke} and 
was able to provide a good simultaneous description of $\gamma\gamma \to \pi^0\pi^0$,
$\pi^+\pi^-$, $\eta \pi^0$, $K^+K^-$ and $K^0\bar{K}^0$ from
threshold up to rather high energies, $s^{1/2}\lesssim 1.5$~GeV. This approach 
was also used in ref.\cite{rocapela} to study the $\eta\to\pi^0\gamma\gamma$ 
decay.

We concentrate here on the dispersive method of 
refs.\cite{penprl,penmorgan,penanegra}. We critically  review and extend it, 
so as we  are able to drastically reduce the uncertainty due to the not-fixed phases 
 above the $K\bar{K}$ threshold of the $I=0$ S-wave $\gamma\gamma\to \pi\pi$
  amplitude. This is accomplished by using 
an $I=0$ S-wave Omn\`es function that is continuous under changes 
in the phase function employed for its evaluation above the $K\bar{K}$ threshold. 
Equivalently,  one can introduce an additional subtraction
in the dispersion relation to evaluate the 
$I=0$  S-wave $\gamma\gamma\to\pi\pi$ amplitude 
 to those  considered in ref.\cite{penprl,penmorgan,penanegra,dh93}. 
This new subtraction constant is fixed by considering simultaneously 
three constraints instead of the two employed
 in the previous references. 
As a result of this much reduced uncertainty, the 
 total cross section $\sigma(\gamma\gamma\to\pi^0\pi^0)$  might  be used 
to distinguish between  different S-wave $I=0$ phase shift 
parameterizations once new more precise experimental data become available. 
 We also  perform calculations of the width 
$\Gamma(\sigma\to\gamma\gamma)$, taking from the literature different 
$\sigma$ resonance parameters, and compare  with the value of ref.\cite{penprl}.
  Other 
papers dedicated to calculate the two photon decay widths of hadronic resonances are
 \cite{rosner}.

The content of the paper is as follows. In section \ref{sec:dis} we
 discuss the dispersive method of
 ref.\cite{penmorgan}  and extend it
  to calculate with higher accuracy the cross section 
 $\gamma\gamma\to \pi^0\pi^0$. The resulting 
 $\sigma(\gamma\gamma\to\pi^0\pi^0)$ and $\Gamma(\sigma\to\gamma\gamma)$ are given in 
 section \ref{sec:resul}. We elaborate our conclusions in section \ref{sec:conclu}.

%------------------------------------------------------------------------------
%-----------------------------------------------------------------------------
\section{Dispersive approach to $\gamma\gamma\to \pi^0\pi^0$}
\label{sec:dis}
\def\theequation{\arabic{section}.\arabic{equation}}
\setcounter{equation}{0}

\begin{figure}[ht]
\psfrag{R}{$P^+$}
\psfrag{Q}{$P^-$}
\centerline{\epsfig{file=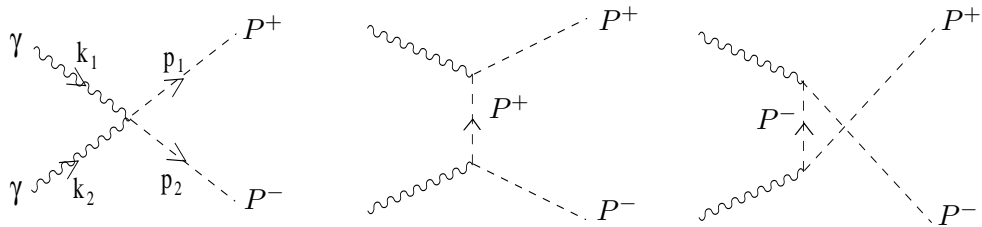,height=1.1in,width=5.0in,angle=0}}
\vspace{0.2cm}
\caption[pilf]{\protect \small
 Born term contribution to $\gamma(k_1) \gamma(k_2)\to P^+(p_1) P^-(p_2)$.  
\label{born}}
\end{figure}

In refs.\cite{penprl,penmorgan,penanegra} an interesting approach was established 
to calculate  in terms of a dispersion relation the 
$\gamma\gamma\to (\pi\pi)_I$ S-wave amplitudes, $F_I(s)$, where the two pions have 
 definite isospin $I$.  Notice that for $\gamma\gamma \to \pi^0\pi^0$, due 
to the null charge of $\pi^0$, there is no Born term, fig.\ref{born}. 
 One then expects, as remarked in ref.\cite{penmorgan},  that only the S-wave would be the important partial wave 
at low energies, $\sqrt{s}\lesssim 0.7$~GeV. For $\gamma\gamma\to \pi^+\pi^-$, 
where there is a Born
 term due to the exchange of charged pions, the D-waves have a relevant contribution 
already at rather low energies due to the smallness of the pion mass. 
 In the following, we shall restrict ourselves to the S-wave contribution to $\gamma\gamma\to
 \pi^0\pi^0$. The explicit  calculation of ref.\cite{oo98} indicates that
 the D-wave contribution at $\sqrt{s}\simeq 0.65$~GeV is smaller than 
 a $10\xpc$ in $\sigma(\gamma\gamma\to \pi^0\pi^0)$,
  and it rapidly decreases for lower energies.

 The function $F_I(s)$ is an analytic function on the complex $s-$plane
  except for two cuts along the
 real axis. The right hand cut happens for $s\geq 4m_\pi^2$, with $m_\pi$ the pion mass,
  and is due to unitarity. The 
 left hand cut, in turn, runs for $s\leq 0$ and is due to unitarity in crossed channels. 
 Let us denote by $L_I(s)$ the complete left hand cut contribution. Then the
 function $F_I(s)-L_I(s)$, by definition, only has right hand cut. 
  Next, refs.\cite{penprl,penmorgan,penanegra}   consider the Omn\`es function
$\omega_I(s)$, 
\be
\omega_I(s)=\exp\left[\frac{s}{\pi}\int_{4m_\pi^2}^\infty \frac{\phi_I(s')}
{s'(s'-s)}ds'\right]~,
\label{omnes}
\ee
with $\phi_I(s)$ the phase of $F_I(s)$ modulo $\pi$, chosen in such a way that 
 $\phi_I(s)$ is {\it continuous} and $\phi_I(4m_\pi^2)=0$. Because of the 
 choice of the phase function $\phi_I(s)$ in eq.(\ref{omnes}), the function 
 $F_I(s)/\omega_I(s)$ has no right hand cut. Then refs.\cite{penprl,penmorgan,penanegra} 
 perform a twice subtracted dispersion relation for $(F_I(s)-L_I(s))/\omega_I(s)$,

 \be
F_I(s)=L_I(s)+a_I \,\omega_I(s)+c_I s\,\omega_I(s)+
\frac{s^2}{\pi}\omega_I(s)\int_{4m_\pi^2}^\infty
\frac{L_I(s')\sin\phi_I(s')}{{s'}^2(s'-s)|\omega_I(s')|}ds'~.
\label{dispen}
\ee
On the other hand, Low's theorem \cite{low} requires $F_I\to B_I(s)$ for $s\to 0$, 
with $B_I$ the Born term contribution, shown in fig.\ref{born}.
 If we write $L_I=B_I+R_I$, with  $R_I\to 0$ for $s\to 0$, as can always be done,
 then Low's theorem  implies also that
 $F_I-L_I\to 0$ for $s\to 0$ and hence $a_I=0$.

For the exotic $I=2$ S-wave one can invoke Watson's final state theorem\footnote{This theorem implies 
that the phase of $F_I(s)$ where there is no inelasticity is the same, modulo $\pi$,
 as the one of the isospin $I$ S-wave $\pi\pi$ elastic strong amplitude.} so that 
$\phi_2(s)=\delta_\pi(s)_2$, the  $I=2$ S-wave $\pi\pi$ phase shifts.  
For $I=0$ the same theorem guarantees that $\phi_0(s)=\delta_\pi(s)_0$ for $s\leq 4m_K^2$, 
with $\delta_\pi(s)_0$ the S-wave $I=0$ $\pi\pi$ phase shifts and $m_K$ the kaon mass. 
Here one neglects the inelasticity due to the $4\pi$ and $6\pi$
states below the two kaon threshold, an accurate assumption 
 as indicated by experiment \cite{hyams,grayer}.
 Above the two kaon threshold, $s_K=4m_K^2$, 
  the phase function $\phi_0(s)$ cannot 
be fixed {\it a priori} due to 
the onset of inelasticity. This is why ref.\cite{penprl} took  for $s>s_K$ that either
$\phi_0(s)\simeq \delta_\pi(s)_0$ or $\phi_0(s)\simeq \delta_\pi(s)_0-\pi$, in order
 to study the size of the  uncertainty induced for low energies. It results, however, that 
 this uncertainty increases dramatically with energy such that for 
 $\sqrt{s}=0.5,~ 0.55,~ 0.6$ and $0.65$~GeV
 it is $20,~45,~92$ and $200\,\xpc$, see fig.3 of ref.\cite{penprl}. 
 
\begin{figure}[ht]
%\psfrag{a}{\begin{tabular}{l} {\small ${\Op}$}\\ {\small Seagull} \end{tabular}}
\centerline{\epsfig{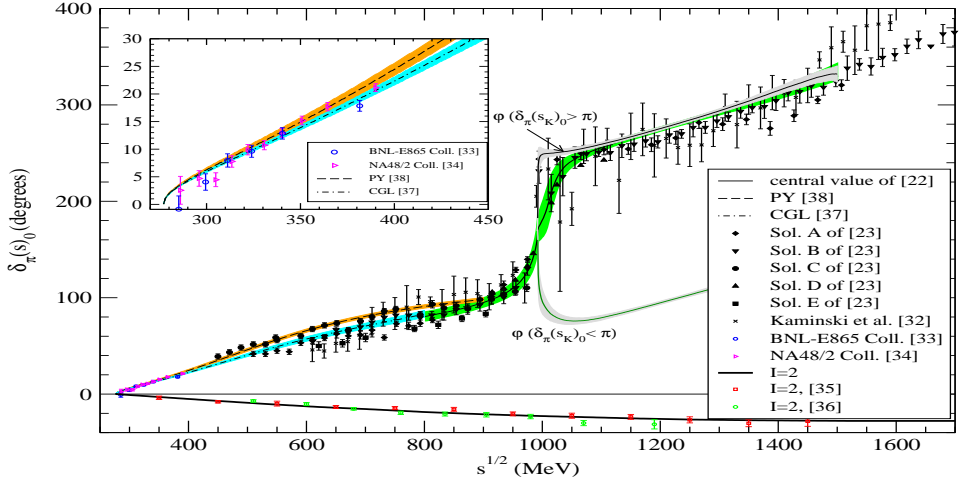}}
\vspace{0.2cm}
\caption[pilf]{\protect \small The phase shifts $\delta_\pi(s)_0$ and 
$\delta_\pi(s)_2$ and  the phase $\varphi(s)$.
 Experimental data are from refs.\cite{kaminski,grayer,bnl,na48} for $I=0$ and 
 refs.\cite{losty,hoog} for $I=2$. The insert is the comparison of 
  CGL \cite{cgl} and PY \cite{py03} with the accurate data 
 from $K_{e4}$ \cite{bnl,na48}.
\label{fig:phases}}
\end{figure}

\begin{figure}[ht]
\centerline{\epsfig{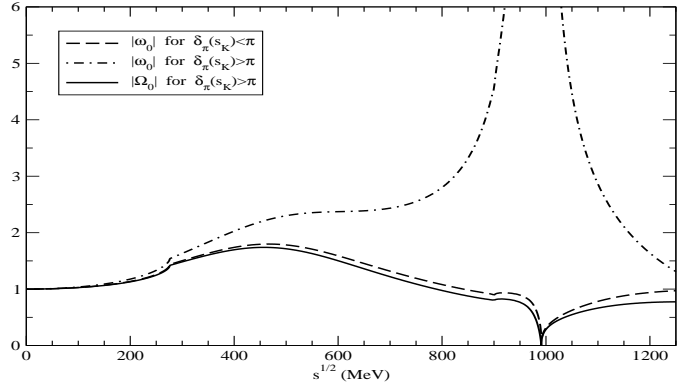}}
\vspace{0.2cm}
\caption[pilf]{\protect \small
 $|\omega_0(s)|$, eq.(\ref{omnes}),  with $\delta_\pi(s_K)_0<\pi$, dashed-line, 
 and $\delta_\pi(s_K)_0>\pi$, dot-dashed line. The solid line is  
  $|\om_0(s)|$, eq.(\ref{om0}), for the latter case. Here $\varphi(s)$ is used 
as  $\phi_0(s)$  in eqs.(\ref{omnes}) and (\ref{om0}) for illustrative purposes.
\label{figomnes}}
\end{figure}

 The reason for this behaviour is the use of the function $\omega_0(s)$ in 
eq.(\ref{dispen}). The $I=0$ S-wave  phase shift $\delta_\pi(s)_0$ has 
a rapid increase by $\pi$ around $s_K=4m_K^2$, 
 due to the narrowness of the $f_0(980)$ resonance on top 
of the $K\bar{K}$ threshold. Let us denote by $\varphi(s)$ the phase of the $\pi\pi\to\pi\pi$ 
 $I=0$ S-wave strong amplitude, modulo $ \pi$, such that it is continuous and $\varphi(4m_\pi^2)=0$. 
 This phase is shown in fig.\ref{fig:phases} together with 
  $ \delta_\pi(s)_0$ and $\delta_\pi(s)_2$. 
  Now, if one uses $\varphi(s)$ instead of $\phi_0(s)$ in eq.(\ref{omnes})
 for illustration,  the function $\omega_0(s)$ is discontinuous 
 in the transition from $\delta_\pi(s_K)_0\to\pi-\epsilon$ to
  $\delta_\pi(s_K)_0\to \pi +\epsilon$, with $ \epsilon\to 0^+$. 
 In the first case $|\omega_0(s)|$ has a zero at $s_K$, while in the latter it becomes $+\infty$.
 This discontinuity is illustrated in fig.\ref{figomnes} by 
 considering the difference between the dot-dashed and dashed lines. 
This discontinuous behaviour of $\omega_0(s)$ under small (even tiny)
 changes of $\delta_\pi(s)_0$ around $s_K$, 
 was the reason for the controversy regarding the value of the pion scalar radius $\la r^2\ra^\pi_s$ 
between \cite{y04,y05,y06} and \cite{accgl05}. This controversy was finally solved in 
 ref.\cite{or07} where it is 
shown that Yndur\'ain's method is compatible with the solutions obtained by solving 
the Muskhelishvili-Omn\`es equations for the scalar form factor \cite{dgl,mous,moo}. 
 The problem arose because  refs.\cite{y04,y05} overlooked the proper solution 
 and stuck to an unstable one. 

   Inelasticity is again small for $1.1 \lesssim \sqrt{s}\lesssim 1.5$~GeV being   
 compatible with zero experimentally \cite{hyams,grayer}. 
 As remarked in refs.\cite{y04,or07}, one can then apply 
 approximately Watson's final state 
 theorem and for $F_0(s)$ this implies that 
  $\phi_0(s)\simeq \delta^{(+)}(s)$ modulo $\pi$. Here  $\delta^{(+)}(s)$ is the eigenphase 
  of the $\pi\pi$, $K\bar{K}$ $I=0$ S-wave S-matrix 
 such that it is continuous and $\delta^{(+)}(s_K)=\delta_\pi(s_K)_0$. In refs.\cite{y05,or07} it is
 shown that $\delta^{(+)}(s)\simeq \delta_\pi(s)_0$ or $\delta_\pi(s)_0-\pi$, depending on whether 
 $\delta_\pi(s_K)_0\geq \pi$ or $<\pi$, respectively. In order to fix the 
 integer factor in front of $\pi$ in $\phi_0(s)\simeq \delta^{(+)}(s)$ modulo $\pi$, 
 one needs to devise an argument to follow the possible 
 trajectories of  $\phi_0(s)$ 
 in the {\it narrow} region $1\lesssim \sqrt{s}\lesssim 1.1$~GeV,  
where inelasticity is not negligible. The remarkable physical effects happening there 
 are the appearance of the 
 $f_0(980)$ resonance on top of the $K\bar{K}$ threshold and the cusp effect of the latter 
 that induces a discontinuity  at $s_K$ in the derivative  of observables, this is clearly visible 
 in  fig.\ref{fig:phases}. 
 Between 1.05 to 1.1~GeV there are no further narrow structures and observables evolve smoothly. 
 Approximately half of the region between 
0.95 and 1.05~GeV is elastic and $\phi_0(s)= \delta_\pi(s)_0$ (Watson's theorem), so that it raises 
 rapidly. Above $2 m_K\simeq 1$~GeV up to 1.05~GeV the function $\phi_0(s)$ can keep increasing with 
 energy, as $\delta_\pi(s)_0$ or $\varphi(s)$ for $\delta_\pi(s_K)_0\geq \pi$, and
this is also always the case for the corresponding phase function of
 the strange scalar form factor of the pion \cite{or07}. It is also the behaviour 
 for $\phi_0(s)$ corresponding to the explicit calculation of ref.\cite{oo98}.  
 The other possibility is a change  of sign in the slope at $s_K$ due to the 
$K\bar{K}$ cusp effect such 
 that $\phi_0(s)$  starts a rapid decrease in 
 energy, like $\varphi(s)$ for $\delta_\pi(s_K)_0<\pi$, fig.\ref{fig:phases}.
  Above $\sqrt{s}=1.05$~GeV,  $\phi_0(s)$ matches smoothly with the 
 behaviour for $\sqrt{s}\gtrsim 1.1$~GeV where it is constraint by Watson's final state theorem. 
   As a result of this matching, for $\sqrt{s}\gtrsim 1$~GeV 
  {\it either}  $\phi_0(s)\simeq \delta_\pi(s)_0$ {\it or}
   $\phi_0(s)\simeq \delta_\pi(s)_0-\pi$, corresponding
 to an increasing or decreasing $\phi_0(s)$  above $s_K$, respectively.  
  There is then left an ambiguity of $\pi$ in $\phi_0(s)$ for $1.5 \hbox{~GeV}\gtrsim \sqrt{s}>\sqrt{s_K}$. Our 
  argument also justifies the similar choice of phases in ref.\cite{penprl} above $s_K$ to estimate
 uncertainties.   Let us define the switch $z$ to characterize the behaviour of $\phi_0(s)$
  for $s>s_K$    such that
  $z=+1$ if $\phi_0(s)$ rises with energy and $z=-1$ if it decreases.
 Above 1.5~GeV the phase function employed has little effect in our energy region and 
  we use the same asymptotic phase function as 
 in ref.\cite{or07}, tending either to $2\pi$ ($z=+1$) or $\pi$ ($z=-1$) for $s\to+\infty$. 
 It allows a large uncertainty of $\simeq 2\pi$ at 
 $\sqrt{s}=1.5$~GeV, that  only shrinks logarithmically for higher energies. This uncertainty 
  is included  in our error analysis. Further details are given in ref.\cite{or07}.
 
 Next, we define, the 
  function $\om_0(s)$, similarly as done in ref.\cite{or07}, 
\be
\om_0(s)=\left(1-\theta(z)\frac{s}{s_1}\right)
\exp\left[ \frac{s}{\pi}\int_{4m_\pi^2}^\infty \frac{\phi_0(s')}{s'(s'-s)} ds'\right]~,
\label{om0}
\ee
where $\theta(z)=+1$ for $z=+1$ and 0 for $z=-1$ and $s_1$ is
 the point at which $\phi_0(s_1)=\pi$. The latter
is the only point where the imaginary part of $\om_0(s)$ vanishes around $s_K$ and this fixes the 
 position of the zero.  Now, we perform the same twice subtracted dispersion
  relation as in eq.(\ref{dispen}) but for $(F_0(s)-L_0(s))/\om_0(s)$~,
\be
F_0(s)=L_0(s)+c_0 s\om_0(s)+
\frac{s^2}{\pi}\om_0(s)\int_{4m_\pi^2}^\infty
\frac{L_0(s')\sin\overline{\phi}_0(s')}{{s'}^2(s'-s)|\om_0(s')|}ds'
+\theta(z)\frac{\omega_0(s)}{\omega_0(s_1)}\frac{s^2}{s_1^2}(F_0(s_1)-L_0(s_1))~.
\label{f0f}
\ee
 
In the previous equation we introduce $\overline{\phi}_0(s)$ that is defined 
 as the phase of $\om_0(s)$.  Let us note that in the case 
 $z=+1$  the phase of $\om_0(s)$ for $s>s_1$ 
 is not $\phi_0(s)$ but $\phi_0(s)-\pi$, 
 due to the factor $1-(s+i\epsilon)/s_1$ in $\Omega_0(s)$, eq.(\ref{om0}). 
  Since $\phi_2(s)$, because of Watson's final state theorem, 
 is given by $\delta_\pi(s)_2$, which is small and smooth \cite{losty,hoog}, fig.\ref{fig:phases}, 
the issue of the discontinuity in $\omega_2(s)$ under changes in parameterizations
 does not rise and we use the dispersion relation in eq.(\ref{dispen}).
 It is worth mentioning that our eq.(\ref{f0f})  for $z=+1$  is equivalent to take 
 a three times subtracted dispersion relation for 
$(F_0(s)-L_0(s))/\omega_0(s)$, two subtractions are taken  at $s=0$ and another one 
at $s_1$. We could have  taken the three subtractions at $s=0$,
 although we find more convenient eq.(\ref{f0f}) which is physically motivated by the use
  of the Omn\`es function eq.(\ref{om0}) that is continuous
   under changes in the parameterization of the $I=0$ 
  S-wave S-matrix. When eq.(\ref{om0}) is used with
  $\varphi(s)$ instead of $\phi_0(s)$ for $\delta_\pi(s_K)_0>\pi$ the solid curve in 
fig.\ref{figomnes} is obtained, which is again close to the dashed line for $\delta_\pi(s_K)_0<\pi$.
  
We denote by $F_N(s)$ the  S-wave $\gamma\gamma\to \pi^0\pi^0$ amplitude and by $F_C(s)$ the
$\gamma\gamma\to\pi^+\pi^-$ one. 
The relations between $F_0$, $F_2$  and $F_N(s)$, $F_C(s)$ in our isospin convention are  
\ba
F_N(s) = -\frac{1}{\sqrt3} F_0 + \sqrt{\frac23} F_2~,~
F_C(s) = -\frac{1}{\sqrt3} F_0 - \sqrt{\frac16} F_2 ~.
\label{fn}
\ea

We have the unknown constants $c_0$, $c_2$  and 
$F_0(s_1)-L_0(s_1)$, the latter for $z=+1$. 
 To determine them we impose: 

\noindent
{\bf 1.} $F_C(s)-B_C(s)$ vanishes linearly in $s$ for $s\to 0$ and we match the
coefficient to the one loop $\chi$PT result \cite{bc88,dhl87}. 
  
\noindent
{\bf 2.} $F_N(s)$  vanishes linearly for $s\to 0$ as well and the coefficient can be 
 obtained again by matching with one loop $\chi$PT \cite{bc88,dhl87}.

\noindent
{\bf 3.} For $I=0$ and $z=+1$ one has still $F_0(s_1)-L_0(s_1)$. The value of this
constant can be restricted taking into account that $F_N(s)$ has an Adler zero, 
due to chiral symmetry. This zero is located at $s_A= m_\pi^2$ in one loop $\chi$PT and 
 moves to $s_A=1.175 \,m_\pi^2$ in two loop $\chi$PT \cite{bgs94}. This implies about a 20$\xpc$
   correction, that prevents us from taking a definite value for $s_A$. In turn, 
 we obtain that the value of the resulting cross section
  $\sigma(\gamma\gamma\to\pi^0\pi^0)$ around the $f_0(980)$ resonance is quite sensitive
  to the position of the Adler zero,  because it controls the  
 size of $F_0(s_1)-L_0(s_1)$. The latter appears in the last term in eq.(\ref{f0f}), the 
 one that dominates $F_0(s)$  
  around the $f_0(980)$ position since $\Omega_0(s_1)=0$. 
 Though the  dispersive method is devised at its best 
for lower energies, it is also clear that it should give at least the proper 
order of magnitude for $\sigma(\gamma\gamma\to \pi^0\pi^0)$ in
  the $f_0(980)$ region.\footnote{E.g., other models
  \cite{oo98,acharev,mennessier} having similar physical mechanisms  describe that energy region very well indeed.} 
  Being conservative,  we shall restrict the values of $F_0(s_1)-L_0(s_1)$ 
 so as the cross section at $s_1$ is less than 10 times the experimental value around 
 the $f_0(980)$ region, $\sigma(\gamma\gamma\to \pi^0\pi^0)<400$~nb.

Regarding $L_I(s)$, it is  expected to be dominated at low energies 
 by the Born term in
isospin $I$ because of the smallness of the pion mass. The Born term originates by the
exchanges in the $t$ and $u$ channels ($\gamma\pi\to\gamma\pi)$ of charged pions, fig.\ref{born}.
 Other crossed exchanges of vector and axial
vector resonances are relatively suppressed for $F_I(s)$ due to the larger masses
 of the members of the $J^{PC}=1^{--}$, $1^{++}$ and $1^{+-}$  multiplets 
 so that their associated left hand cut singularities  are further away. 
 Among them, the $1^{++}$ axial vector exchange contributions are the dominant ones 
in $\gamma \pi^\pm\to \gamma\pi^\pm$ and already appear at the one loop level in $\chi$PT. 
The $1^{--}$ and $1^{+-}$ exchanges start one order higher. 
 As already remarked in ref.\cite{dh93}, the authors of 
refs.\cite{penmorgan,penanegra}, and later on also  ref.\cite{penprl}, 
  overlooked the axial vector exchange contributions altogether 
 and hence they are missing an essential part in the
 study of the low energy $\gamma\gamma\to\pi^0\pi^0$. 
Indeed, once the $1^{++}$ axial vector
 exchange contributions are considered the cross section
  in the latter references would increase significantly at low energies. For instance, at 
  $\sqrt{s}=0.5$~GeV one has more than a $20\xpc$ increase compared to the case in which only the Born
  term and the $1^{--}$ vector resonances exchanges are considered. 
  At low energies the influence of the $1^{+-}$ axial
  vector nonet is much smaller than that of the $1^{++}$  and  $1^{--}$ multiplets.   
  In the following, $L_I(s)$ is modelled by the Born terms and the crossed exchanges of 
the $1^{++}$, $1^{--}$,  and $1^{+-}$ resonance multiplets, evaluated from chiral Lagrangians.
 Explicit expressions for the distinct contributions to $L_I(s)$ will be given elsewhere
  \cite{orslong}.

%%%%%%%%%%%%%%%%%%%%%%%%%%%%%%%%%%%%%%%%%%%%%%%%%%%%%%%%%%%%%%%%%%%%%%%%%%%%%
%%%%%%%%%%%%%%%%%%%%%%%%%%%%%%%%%%%%%%%%%%%%%%%%%%%%%%%%%%%%%%%%%%%%%%%%%%%%%
\section{Results}
\label{sec:resul}

\begin{figure}[ht]
%\psfrag{a}{\begin{tabular}{l} {\small ${\Op}$}\\ {\small Seagull} \end{tabular}}
\centerline{\epsfig{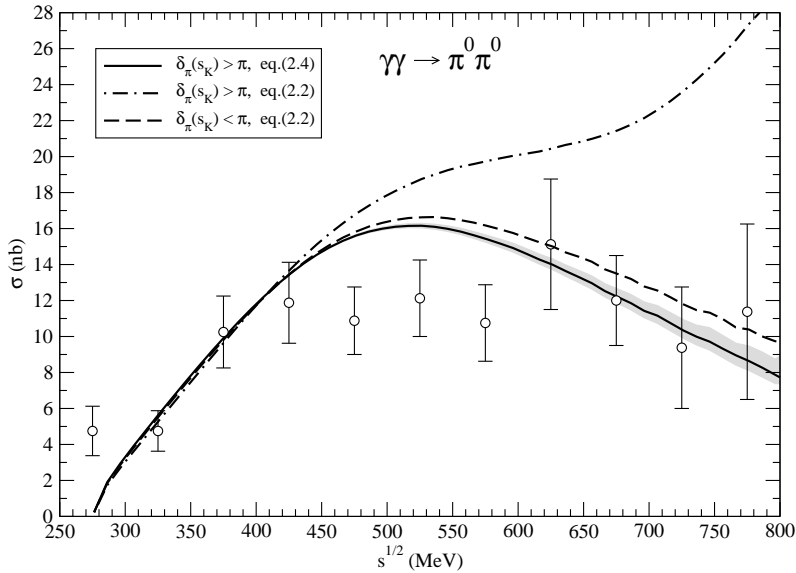}}
\vspace{0.2cm}
\caption[pilf]{\protect \small
 The solid line corresponds 
to $z=+1$ and the error band is the uncertainty by requiring 
that $\sigma(\gamma\gamma\to\pi^0\pi^0)<400$~nb at $s_1$. This line should be
compared with the dot-dashed one that would result from the formalism of ref.\cite{penprl}, 
including axial vector exchanges. 
 Finally, the dashed line corresponds to $z=-1$. \label{fig:updown}}
\end{figure}

In this section we show the results that follow by the use of eq.(\ref{f0f}) for $I=0$ and 
eq.(\ref{dispen}) for $I=2$. Since the main contribution in the low energy region 
to $\om_0(s)$ and $\omega_2(s)$  comes from
the low energy $\pi\pi$ phase shifts, one needs to be as precise as
possible for low energy $\pi\pi$ scattering data. The small $I=2$ S-wave $\pi\pi$ phase
shifts, which induce small final state interaction corrections anyhow,
can be parameterized in simple terms and our fit compared to data can be seen in 
 fig.\ref{fig:phases}. For the $I=0$ S-wave $\pi\pi$,  we take the
parameterizations of ref.\cite{cgl} (CGL) and ref.\cite{py03} (PY). 
Both  agree with data from $K_{e4}$ decays 
\cite{bnl,na48} and span to a large extend the band of theoretical uncertainties in 
the $I=0$ S-wave $\pi\pi$ phase shifts
 \cite{hyams,grayer,kaminski}.  PY, similarly to refs.\cite{npa,nd,iamcc}, runs 
 through the higher values of $\delta_\pi(s)_0$, while CGL
  does through lower values, see fig.\ref{fig:phases}. 
 We shall use CGL up to
 0.8~GeV, since this is the upper limit of its analysis, and the
K-matrix of Hyams {\it et al.} \cite{hyams} above that energy. The latter corresponds to the 
energy dependent analysis of the experimental data of the same reference. 
On the other hand, PY is used up to 0.9~GeV, since at that
 energy this parameterization agrees well inside errors with
\cite{hyams}, and above 0.9~GeV the K-matrix 
of ref.\cite{hyams} is taken. Given the input functions $\phi_I(s)$ and $L_I(s)$, 
the constants $c_2$, $c_0$ and $F_0(s_1)-L_0(s_1)$ can be fixed by the three conditions 
 explained at the end of section \ref{sec:dis}. The 
$\gamma\gamma\to\pi^0\pi^0$ S-wave amplitude $F_N(s)$, eq.(\ref{fn}), 
can then be calculated and the total cross section is given by
$\sigma(\gamma\gamma\to\pi^0\pi^0)=\frac{\beta}{64\pi s}|F_N(s)|^2~,$
with $\beta(s)=\sqrt{1-4m_\pi^2/s}$. 
 We show in fig.\ref{fig:updown} the drastic reduction in the uncertainty of the cross section
due to the variation of $\phi_0(s)$ above $s_K$ as commented in the previous section. 
 For $z=+1$ one has the solid line while for $z=-1$ the dashed line results, both are
 very close. This should be compared with the dot-dashed line that is obtained from the approach 
  of refs.\cite{penprl,penmorgan,penanegra}. The uncertainty now, by employing 
 eq.(\ref{f0f}) instead of eq.(\ref{dispen}) for $I=0$, is drastically reduced. This improvement also
 implies  that 
 our results can be compared with data for $\sqrt{s}\gtrsim0.5$~GeV.
   We also show by the gray band around the solid line the mild influence in our
 calculations of the uncertainty in the location of the Adler zero, restricted so that
$\sigma(\gamma\gamma\to \pi^0\pi^0)<400$~nb at the $f_0(980)$ region (experiment is $\simeq
40$~nb). 

\begin{figure}[ht]
%\psfrag{CGL}{CGL}
%\psfrag{PY}{PY}
\psfrag{ChPT at one loop}{{\small $\chi$PT to one loop}}
\psfrag{ChPT at two loops}{{\small $\chi$PT to two loops}}
\centerline{\epsfig{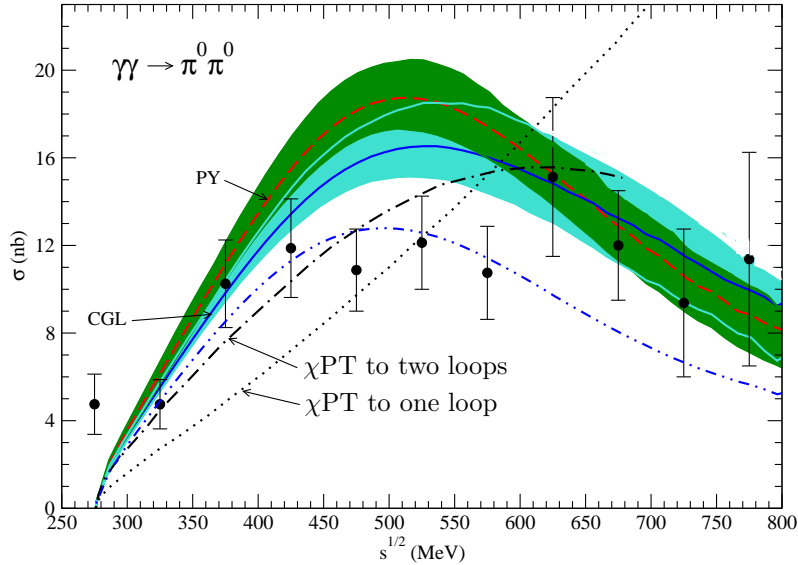}}
\vspace{0.2cm}
\caption[pilf]{\protect \small
Final results for the $\gamma\gamma\to\pi^0\pi^0$ cross
section. Experimental data are from the Crystal Ball Coll. \cite{crystal},
 scaled by $1/0.8$, as   $|\cos\theta|<0.8$ is measured and S-wave dominates. 
 The solid line corresponds to CGL and the dashed one to PY. 
The dot-dot-dashed line results after removing the axial vector exchange contributions.
 The band along each line represents the theoretical uncertainty. The dotted line 
 is the one loop $\chi$PT result \cite{bc88,dhl87} and the dot-dashed one the two loop calculation  
 \cite{bgs94}. 
\label{fig:errorband}}
\end{figure}

Our final $\sigma(\gamma\gamma\to\pi^0 \pi^0)$ is shown in fig.\ref{fig:errorband}. We give 
the corresponding results for CGL(solid) and PY(dashed), 
 where the band around every line 
stems from the uncertainties in our approach, which comprise:  the errors in the
 Hyams {\it et al.} \cite{hyams}, CGL and PY parameterizations 
 (those of the last two indeed dominate the width of the bands),
  to use  either $\phi_0(s)\simeq \delta_\pi(s)_0$
  or  $\delta_\pi(s)_0-\pi$ for $s>s_K$, the uncertainty in the asymptotic phase
  and to restrict $\sigma(\gamma\gamma\to \pi^0\pi^0)<400$~nb in the 
  $f_0(980)$ region for $z=+1$. 
 On top of that, we evaluate  the conditions {\bf 1} and {\bf 2} above from 
 the expressions given by one loop $\chi$PT 
  either by employing $f_\pi=92.4$~MeV or $f\simeq 0.94 f_\pi$, where 
the former is the pion decay constant and the latter is the same but in the $SU(2)$ chiral limit 
 \cite{glchpt}. 
 This amounts to around a $12\xpc$ of uncertainty in the evaluation of $c_0$ and $c_2$, due
 to the square dependence on $f_\pi$. 
Note that both choices, $f_\pi$ or $f$, are consistent with the 
 precision of the one loop calculation and the variation in
the results is an estimate for higher order corrections.
 However, the error induced in $\sigma(\gamma\gamma\to
 \pi^0\pi^0)$ is much smaller than that from the other sources of uncertainty
  and can be neglected when  added in quadrature. 

  In the same figure the dotted line corresponds to one loop CHPT
 \cite{bc88,dhl87} and the dot-dashed one to the two loop result \cite{bgs94,gis05}. The latter
  is closer to our results but still one observes that the ${\cal O}(p^8)$ corrections
 would be sizable. 
 It is worth stressing that if the axial vector exchanges were removed, as in
 refs.\cite{penprl,penanegra}, then our curves would be smaller. This corresponds to the 
dot-dot-dashed line in fig.\ref{fig:errorband} which is very close to that of 
 ref.\cite{penprl} when employing $\phi_0(s)\simeq \delta_\pi(s)_0-\pi$
for $s>s_K$. This curve is evaluated making use of CGL and ref.\cite{hyams}. The three experimental points \cite{crystal}
 in the region $0.45-0.6$~GeV agree well with this curve.
However, once the axial vector are included the curve rises. These three points lie around 1.5
 sigmas below the CGL result band, and by more than two sigmas below 
 the PY one. This clearly shows that more precise experimental data on 
 $\gamma\gamma\to \pi^0\pi^0$  could  be used to distinguish between  
 different  S-wave  parameterizations. 
 In turn, the next three experimental $\sigma(\gamma\gamma\to\pi^0\pi^0)$ points, 
 those lying between $0.6-0.75$~GeV, agree better when 
 the axial vector resonance contributions are taken into account, as one should do. 
 As a result of this discussion, more precise experimental data for
 $\gamma\gamma\to\pi^0\pi^0$ are called for.

In terms of the calculated $F_N(s)$ one can evaluate the $\sigma$ coupling to
$\gamma\gamma$, called $g_{\sigma\gamma\gamma}$. 
  The dispersion relation to calculate $F_0(s)$ is only valid
on the first Riemann sheet. If evaluated on the second Riemann sheet there
would be an extra term due to the $\sigma$ pole. 
However, the relation between $F_0(s)$ and
$\widetilde{F}_0(s)$, the latter on the second sheet, can be easily
established by using unitarity above the $\pi\pi$ threshold,
\be
F_0(s+i\epsilon)-F_0(s-i\epsilon)=-2i F_0(s+i\epsilon)\rho(s+i\epsilon)T_{II}^0(s-i\epsilon)~,
\label{f01}
\ee
 with $4m_\pi^2\leq s \leq 4m_K^2$, $\rho(s)=\beta(s)/16\pi$ and $\epsilon\to0^+$. In the equation above $T_{II}^0(s)$ is the 
 $I=0$ S-wave $\pi\pi$ elastic amplitude on the second Riemann sheet and $T_{I}^0(s)$ is the one on 
 the  physical Riemann sheet. 
  Due to  continuity when changing from one sheet to the other,
 $
 F_0(s-i\epsilon)=\widetilde{F}_0(s+i\epsilon)~,~T_I^{I=0}(s-i\epsilon)={T}_{II}^{I=0}(s+i\epsilon)~.
$ Then, eq.(\ref{f01}) can be rewritten as
\be
\widetilde{F}_0(s)=F_0(s)\left(1+2i \rho(s)T_{II}^{I=0}(s)\right)~.
\label{analytic}
\ee
 Around the $\sigma$ pole, $s_\sigma$, 
\be
T_{II}^{I=0}=\frac{g_{\sigma \pi\pi}^2}{s_\sigma-s}~,~\widetilde{F}_0(s)=\sqrt{2}\,\frac{g_{\sigma
\gamma\gamma}g_{\sigma\pi\pi}}{s_\sigma-s}~,
\label{sqrt2}
\ee
with $g_{\sigma\pi\pi}$ the $\sigma$ coupling to two pions 
such that $\Gamma=|g_{\sigma\pi\pi}|^2 \beta/16\pi M$, for a narrow enough
scalar resonance of mass $M$. 
Notice as well 
the $\sqrt{2}$ factor in $\widetilde{F}_0(s)$ to match with the 
$g_{\sigma\pi\pi}$ normalization used (the so called unitary normalization \cite{npa,nd,iamcc}). 
Then from eqs.(\ref{analytic}) and (\ref{sqrt2}) it follows that
\be
\frac{g_{\sigma\gamma\gamma}^2}{g_{\sigma\pi\pi}^2}=-\frac{1}{2}\left(\frac{\beta(s_\sigma)}{8\pi}\right)^2 
F_0(s_\sigma)^2~,
\label{12fac}
\ee
Let us stress that this equation gives the ratio between the residua of the S-wave $I=0$ 
$\gamma\gamma\to\pi\pi$ and $\pi\pi\to \pi\pi$ amplitudes at the $\sigma$ pole position. 

In order to derive specific numbers for the previous ratio in terms of our dispersive 
approach one needs to introduce $s_\sigma$. We  take two different values for $s_\sigma=(M_\sigma-i\,
\Gamma_\sigma/2)^2$. From  the studies of Unitary $\chi$PT \cite{npa,nd,iamcc,mixing} one has 
 $M_\sigma$ and $\Gamma_\sigma$ around the interval 425-440 MeV. The other values that we will
 use are from ref.~\cite{caprini}, $M_\sigma^{ccl}=
 441^{+16}_{-8}$~MeV and $\Gamma_\sigma^{ccl}=544^{+18}_{-25}$~MeV, 
 where the superscript $ccl$ indicates, in the following, values
 that employ the $\sigma$ pole position of ref.\cite{caprini}. 
 The corresponding  ratios of the residua given in eq.(\ref{12fac}) are:
\ba
\left|\frac{g_{\sigma\gamma\gamma}}{g_{\sigma\pi\pi}}\right|&=&(2.10\pm 0.25)\times 10^{-3}~,
~s_\sigma \hbox{~from ref.\cite{mixing}}~,\nn\\
\left|\frac{g_{\sigma\gamma\gamma}}{g_{\sigma\pi\pi}}\right|&=&(2.06\pm 0.14)\times 10^{-3}~,
~s_\sigma \hbox{~from ref.\cite{caprini}}~.
\label{residuarati}
\ea
Both numbers are very similar despite that the imaginary parts of 
the two $s_\sigma^{1/2}$ differ by  $\sim 20\%$. The result of \cite{penprl}, 
with which we shall compare our results later, corresponds to the ratio in
eq.(\ref{residuarati}) being $20\%$ bigger at $(2.53\pm 0.09)\times 10^{-3}$ with $s_\sigma$ 
of ref.\cite{caprini}.  

These ratios of residua at the $\sigma$ pole position 
 are the well defined predictions that follow from our improved dispersive treatment of 
 $\gamma\gamma\to(\pi\pi)_I$. However,  
  the radiative width to $\gamma\gamma$ for a wide resonance like the $\sigma$, 
  though more intuitive, has experimental 
  determinations that are parameterization dependent. This is  
  due to the non-trivial interplay between background and the broad resonant 
  signal. An unambiguous
  definition is then required \cite{penprl,penanegra}. We
  employ, as in ref.\cite{penprl}, the standard narrow resonance width formula in terms of
  $g_{\sigma\gamma\gamma}$ determined from eq.(\ref{12fac}) by calculating the residue at $s_\sigma$,
\be
\Gamma(\sigma\to \gamma\gamma)=\frac{|g_{\sigma\gamma\gamma}|^2}{16\pi M_\sigma}~.
\label{sigmafoton}
\ee
 Nevertheless, the determinations of the radiative widths  from this expression and those 
 from common experimental analyses can differ substantially.
 The following example makes this point clear. 

From ref.\cite{mixing} one obtains $|g_{\sigma\pi\pi}|=2.97-3.01$~GeV, corresponding to the square root 
of the residua of the $I=0$ S-wave $\pi\pi$ amplitude, as in eq.(\ref{sqrt2}). 
If similarly to eq.(\ref{sigmafoton}), one  uses the formula, 
\be
\Gamma_\sigma=\frac{|g_{\sigma\pi\pi}|^2 \beta(M_\sigma)}{16 \pi M_\sigma}~,
\label{gammalower}
\ee
the resulting width lies in the range $309-319$~MeV, that is around a 30$\%$ smaller than 
$\Gamma_\sigma\simeq 430$~MeV from the pole position of ref.\cite{mixing}. This is due 
to the large width of the $\sigma$ meson which makes the $|g_{\sigma\pi\pi}|$ extracted 
from the residue of $T_{II}^{I=0}$, eq.(\ref{sqrt2}), be smaller by around a $15\%$ than 
  the value needed in eq.(\ref{gammalower}) to obtain $\Gamma_\sigma\simeq 430$~MeV. 
 Similar effects are then also expected in order to extract $\Gamma(\sigma\to\gamma\gamma)$ 
 from the eq.(\ref{sigmafoton}). 
Equations similar to this are usually employed in phenomenological fits to data, e.g. see
ref.\cite{mori}, but with $|g_{\sigma\gamma\gamma}|$ determined along the real axis.
As a result of this discussion, one should allow 
a $(20-30)\%$ variation between the results obtained from eq.(\ref{sigmafoton}) and those 
from standard experimental analyses that still could 
deliver a $\gamma\gamma\to\pi\pi$ amplitude 
in agreement with our more theoretical treatment for physical values of $s$.

We shall employ the following values for $|g_{\sigma\pi\pi}|$. 
First we take $|g_{\sigma\pi\pi}|= 2.97-3.01$~GeV
\cite{npa,nd,iamcc,mixing}. With this value the resulting two photon 
width from eqs.(\ref{residuarati}) and (\ref{sigmafoton}) is
\be
\Gamma(\sigma\to\gamma\gamma)=(1.8\pm 0.4) \hbox{~KeV}~.
\label{gammamix}\ee
We  also consider a larger value for $|g_{\sigma\pi\pi}|$ since $\Gamma_\sigma^{ccl}$ \cite{caprini} is 
larger by a factor $\sim 1.3$ than $\Gamma_\sigma$ from ref.\cite{mixing}. 
 One value is
\be
|{g_{\sigma\pi\pi}|^{ccl}}_{(1)}\simeq 
|g_{\sigma\pi\pi}| \left(\frac{\Gamma^{ccl}(\sigma\to\pi\pi)}{\Gamma(\sigma\to\pi\pi)}\right)^\frac12
=(1.127\pm 0.022)\,|g_{\sigma\pi\pi}|=(3.35\pm 0.08) \hbox{~GeV}~.
\label{value1}\ee 
This corresponds to the scenario discussed previously in eq.(\ref{gammalower}) with a value 
  15$\%$ lower than  
\be
{|g_{\sigma\pi\pi}|^{ccl}}_{(2)}=
\left(
\frac{16 \pi M_\sigma \Gamma^{ccl}_\sigma}{\beta(M_\sigma)}\right)^{1/2}= (3.93\pm 0.08)\hbox{~GeV}~,
\label{value2}
\ee
obtained by reproducing $\Gamma_\sigma^{ccl}$ from the pole position using eq.(\ref{gammalower}). 
 If  we evaluate with these couplings 
 the $\sigma\to\gamma\gamma$ width one obtains from
eqs.(\ref{residuarati}) and (\ref{sigmafoton}), respectively,
\ba 
\Gamma^{ccl}_{(1)}(\sigma\to \gamma\gamma)&=&~(2.1\pm 0.3) \hbox{~KeV}~,\nn\\
 \Gamma^{ccl}_{(2)}(\sigma\to\gamma\gamma)&=&~(3.0\pm 0.3) \hbox{~KeV}~.
\label{twoccl}\ea 
 
Recently, ref.\cite{penprl} calculated a value $\Gamma(\sigma\to\gamma\gamma)=(4.09\pm0.29)~\textrm{KeV}$ 
also employing $s_\sigma$ from ref.\cite{caprini}. 
This value is  larger than $\Gamma^{ccl}_{(2)}(\sigma\to\gamma\gamma)$ in the previous equation, 
despite that 
$|g_{\sigma\pi\pi}|$ there used is 3.86 GeV, very close to $|g_{\sigma\pi\pi}|^{ccl}_{(2)}$. 
 It is worth stressing that both our eq.(\ref{sigmafoton}) and eq.(7) of ref.\cite{penprl} 
  are equivalent for calculating $\Gamma(\sigma\to \gamma\gamma)$, except for an extra factor 
 $|\beta(s_\sigma)|\sim 0.95$ 
 in  ref.\cite{penprl}. Of course, they are written in a different notation.\footnote{We
 want to thank M.R. Pennington for a detailed comparison of his results with ours and 
 interesting discussions.}
 The reason for this remaining difference is two fold.  
As already mentioned above, ref.\cite{penprl} does not include axial vector exchanges in 
 evaluating $\gamma\gamma\to(\pi\pi)_I$. It is this omission that accounts for half 
 of the $20\%$ difference in the ratio of residua, eq.(\ref{residuarati}), mentioned above. 
 The other $10\%$ comes from improvements delivered by our extra subtraction and our slightly
 different inputs. Using the same value for $|g_{\sigma \pi\pi}^{ccl}|$ as in 
 \cite{penprl}, our resulting value for $\Gamma(\sigma\to\gamma\gamma)$ would be around 
 a $40\%$ smaller (as in eq.(\ref{twoccl})) than that in \cite{penprl}.

 As a summary of the $\sigma\to\gamma\gamma$ considerations, 
 from our dispersive approach and $s_\sigma$ of
 refs.\cite{mixing,caprini} we obtain a  value for the ratio of the residua
 $|g_{\sigma\gamma\gamma}/g_{\sigma\pi\pi}|\sim (2.1\pm 0.25)\times 10^{-3}$. This number follows 
 unambiguously from our study. Other more intuitive, but convention dependent quantities, like the 
 $\sigma\to\gamma\gamma$ width calculated from eq.(\ref{sigmafoton}), 
 are less well determined.  These depend critically on the input value 
 for $|g_{\sigma\pi\pi}|^2$ and $s_\sigma$, though they are not required in our dispersive study of
 $\gamma\gamma\to\pi^0\pi^0$.  We then determine the values: 
 i) $\Gamma(\sigma\to\gamma\gamma)=(1.8\pm 0.4)$~KeV
 with $s_\sigma$ and  $|g_{\sigma\pi\pi}|\sim 3$~GeV from ref.\cite{mixing}; ii) 
  $\Gamma^{ccl}_{(1)}(\sigma\to\gamma\gamma)=(2.1\pm 0.3)$~KeV and 
  $\Gamma^{ccl}_{(2)}(\sigma\to\gamma\gamma)=(3.0\pm 0.3)$~KeV which 
 come by considering $s_\sigma$ of ref.\cite{caprini} 
 with an estimated $|g_{\sigma\pi\pi}|=3.4$ and $3.9$~GeV, respectively.  
 Other values could be obtained 
 from eq.(\ref{sigmafoton}) by plugging
 different $s_\sigma$ and $|g_{\sigma\pi\pi}|$ in eq.(\ref{12fac}) 
in order to estimate $|g_{\sigma\gamma\gamma}|$.
 One should require that these values are provided 
 from a $\pi\pi$ S-wave $I=0$ strong amplitude in agreement with the experimental phase shifts,
 see fig.\ref{fig:phases}.

%%%%%%%%%%%%%%%%%%%%%%%%%%%%%%%%%%%%%%%%%%%%%%%%%%%%%%%%%%%%%%%%%%%%%%%%%%%%%%%%%%
%%%%%%%%%%%%%%%%%%%%%%%%%%%%%%%%%%%%%%%%%%%%%%%%%%%%%%%%%%%%%%%%%%%%%%%%%%%%%%%%%
\section{Conclusions}
\label{sec:conclu}

We have undertaken a dispersive study of the $\gamma\gamma\to \pi^0\pi^0$ reaction. 
Our approach is based on that of refs.\cite{penmorgan,penanegra,penprl} but using 
a better behaved Omn\`es function for the $I=0$ S-wave $\pi\pi$ channel. As a result,
  we have been able to reduce drastically  
 the uncertainty regarding the $\phi_0(s)$ used in this
Omn\`es function above the $K\bar{K}$ threshold. 
 Our improvement is equivalent to take three subtractions 
 instead of the two  originally proposed in refs.\cite{penmorgan,penanegra,penprl}. 
 We have then used two low energy conditions and a third constraint in the form of a bound on the $f_0(980)$ region 
 so as to fix the three subtraction constants. This has allowed
  us to present more accurate results, which 
 might be used to discriminate between different 
 $\pi\pi$ $I=0$ S-wave parametrizations, once more precise 
 data on  $\sigma(\gamma\gamma\to \pi^0\pi^0)$ become available. Further improvements at the theoretical level
 rest on a more precise determination of $\phi_0(s)$  above $s_K$ and 
a more systematic calculation of $L_I(s)$.

 We have calculated the ratio of the residua 
 $|g_{\sigma \gamma\gamma}/g_{\sigma\pi\pi}|=(2.1\pm 0.25)\times 10^{-3}$
  with $s_\sigma$ from
 refs.\cite{mixing,caprini}.  The $\sigma$ width to $\gamma\gamma$ was also studied   
 and we stressed its dependence on the $s_\sigma$ and $|g_{\sigma\pi\pi}|^2$ 
employed, not used in our dispersive study of $\gamma\gamma\to\pi^0\pi^0$.  
 One value obtained is $\Gamma(\sigma\to\gamma\gamma)=(1.8\pm 0.4)$~KeV with $s_\sigma$
and $|g_{\sigma\pi\pi}|\simeq 3$~GeV from ref.\cite{mixing}. 
The others values take $s_\sigma$ as given in ref.\cite{caprini} with 
$\Gamma(\sigma\to\gamma\gamma)=(2.1\pm 0.3)$~KeV for $|g_{\sigma\pi\pi}|=3.4$~GeV, and 
 $\Gamma(\sigma\to\gamma\gamma)=(3.0\pm 0.3)$~KeV for 
 $|g_{\sigma\pi\pi}|=3.9$~GeV. 
 The last two numbers for $\Gamma^{ccl}(\sigma\to\gamma\gamma)$  tell us 
that the uncertainties in its calculation are still rather large 
and a further improvement requires 
to know precisely $|g_{\sigma\pi\pi}^{ccl}|$ from ref.\cite{cgl}. 
 
\section*{Acknowledgements}
We would like to thank E.~Oset for  useful communications.
 This work has been supported in part by the MEC (Spain) and FEDER (EC) Grants
  FPA2004-03470 and Fis2006-03438,  the 
  Fundaci\'on  S\'eneca (Murcia) grant Ref. 02975/PI/05, the European Commission
(EC) RTN Network EURIDICE  Contract No. HPRN-CT2002-00311 and the HadronPhysics I3
Project (EC)  Contract No RII3-CT-2004-506078. C.S. acknowledges 
the Fundaci\'on S\'eneca by funding his stay at  the Departamento de F\'{\i}sica de la 
Universidad de Murcia,  and the latter by its warm hospitality. 

\end{document}